\documentstyle[12pt]{article}
\begin{document}

\overfullrule 0 mm
\language 0
\centerline { \bf {ABSENCE OF RADIATION REACTION }}
\centerline{\bf{FOR AN EXTENDED  PARTICLE}}
\centerline { \bf { IN CLASSICAL 
ELECTRODYNAMICS}}
\vskip 0.3 cm
\centerline {\bf{  Alexander A. Vlasov}} 
\vskip 0.3 cm
\centerline {{ High Energy and Quantum Theory}} \centerline {{  
Department of Physics}} \centerline {{  Moscow State University}} 
\centerline {{ Moscow, 119899}} \centerline {{  Russia}} \vskip 0.3 
cm

{\it There are known problems with the standard Lorentz-Dirac 
description of radiation reaction in classical electrodynamics. 
The model of extended in one dimension particle is proposed 
and is shown that for this model there is no total change in 
particle momentum due to radiation reaction.}

03.50.De
 \vskip 0.3 cm

There are known problems with the standard Lorentz-Dirac description 
of radiation reaction in classical electrodynamics 
(mass renormalization and its nonuniqueness, preacceleration, 
runaway solutions and so on, see, for ex., [1-7]).

In the literature one can find the opinion that some of that 
problems are connected with the "point" description of moving 
charged particle and the hope that the macroscopic model of extended 
particle can get rid of them (see, for ex., [8-10]).

In this article for a 
extended in one dimension charged particle we  deduce the 
equation of motion   and 
discuss some  properties of particle motion along one axis.

Let us take the hydrodynamic model of an extended particle.

Then the particle is described by the mass density $m \cdot f(t,x)$, 
charge density  $\rho(t,x)$ and current density $j(t,x)$, obeying 
 the continuity equations $$m{\partial f \over \partial 
 t}+m{\partial (v\cdot f) \over \partial x}=0;$$ 
 $${\partial \rho \over \partial t}+{\partial j \over 
\partial x}=0 \eqno (1)$$ here $v=v(t,x)$ is the hydrodynamic 
velocity of moving extended particle.

Let the particle move under the external force $F(t,x)$ 
along x-axis.  Then the relativistic equation of its motion reads 
(we choose the units $c=1$):  $$m\int dx f(t,x)\left( {\partial  
\over \partial t}+v{\partial \over \partial x} \right) u =\int dx 
\rho(t,x)E(t,x) + \int dx f(t,x)F(t,x) \eqno (2)$$ here $v=v(t,x)$, 
$u=u(t,x)= v/\sqrt{1-v^2}$, $E(t,x)$ -is the electric field, 
produced by moving particle (the Lorentz force is absent in 
one-dimension case under consideration and internal forces give 
zero contribution to total force):

$$E= - {\partial \phi \over \partial x} -{\partial A \over 
\partial t}  \eqno(3)$$
and electromagnetic potentials $\phi$ and $A$ are
$$\phi(t,x)=\int dx'dt'{\rho(t',x') \over 
|x-x'|}(a\delta_1+b\delta_2),$$
$$A(t,x)=\int dx'dt'{j(t',x') \over 
|x-x'|}(a\delta_1+b\delta_2),\eqno(4)$$
with retarded and advanced delta-functions
$$\delta_1= \delta(t'-t+|x-x'|),\ \ \delta_2= \delta(t'-t-|x-x'|) $$
and $a,b$ -  constants.

Substitution of (4) in (3) and integration by parts with the help of 
eq. (1) (taking zero values for integrals of exact integrands in 
$x'$, i.e. $\int dx' {\partial \over \partial x'} 
(\rho\cdot\cdot\cdot)=0$, yields $$E(t,x)=\int {dx'dt' \over 
|x-x'|^2}\left(\rho(t',x') {x-x'\over 
|x-x'|}(a\delta_1+b\delta_2)+j(t',x')(a\delta_1-b\delta_2)\right) 
\eqno(5)$$
Similar integration by parts for LHS of (2) gives the common result 
$$LHS={dP \over dt},\ \ P=P(t)=m\int dx f(t,x) u(t,x) \eqno(6)$$
here $P$ - the particle momentum.
Thus the eq. of motion reads
$${dP \over dt} =F_{self} +F_{ext},$$
$$F_{self}=\int dx \rho(t,x) E(t,x),\ \ F_{ext}=\int dx f(t,x) 
F(t,x) \eqno(7)$$ This eq. of motion has no second derivative of 
particle velocity; also there is no need in mass renormalization. 
 
If the extended particle is compact, we can use in (5,7) the 
standard expansion in powers of $|x-x'|$ (see, for ex.,[2,3]):
$$\delta (t'-t+\epsilon |x-x'|)=\sum\limits_{n=0}^{\infty} 
{\epsilon ^n |x-x'|^n\over n! }{\partial^n \over (\partial 
t')^n}\delta(t'-t)$$
with $\epsilon=\pm 1$.

Thus in nonrelativistic case we get the known result:
$$F_{self}=-(a+b)\int {dx'dt' \over |x-x'|}\rho(t,x')\rho(t,x)
{\partial v(t,x') \over \partial t}+$$
$${2\over3}(a-b)\int dx'dt'\rho(t,x')\rho(t,x)
{\partial^2 v(t,x') \over (\partial t)^2}$$

Now consider the problem of runaway solutions and self-interaction.

The total 
change in particle momentum is $$\Delta P= P(\infty)-P(-\infty)=\int 
dt {dP \over dt}$$ and the change in particle momentum due to its 
self-interaction is $$\Delta P_{self}=\int dt F_{self} \eqno (8)$$ 
Thus
$$ \Delta P =\Delta P_{self}+\int dt F_{ext}    \eqno(9)$$
Substitution of (5) into (7-8) gives
 $${ d  P_{self} \over dt}=F_{ext}=\int 
dt'dxdx' {\rho(t,x) \over |x-x'|^2}\cdot$$
$$ \left(\rho(t',x') 
{x-x'\over 
|x-x'|}(a\delta_1+b\delta_2)+j(t',x')(a\delta_1-b\delta_2)\right) 
\eqno(10)$$
$$\Delta P_{self} =\int dt \left[RHS\ \  of\ \  (10)\right] 
\eqno(11)$$
The solution of (1) we can write as
$$\rho(t,x)= {\partial \Phi(t,x) \over \partial x},\ \ j(t,x)= 
-{\partial \Phi(t,x) \over \partial t} \eqno(12)$$
Then integration by parts in (10,11) with the help of (12) gives    
the following result:  $$\Delta P_{self}=\int 
dtdt'dxdx'\Phi(t,x)\Phi(t',x'){x-x'\over |x-x'|^4}\cdot$$ 
$$\left[-6(a\delta_1+b\delta_2) 
+6|x-x'|\left(a{\partial \delta_1 \over \partial t'}+b{\partial 
\delta_2 \over \partial t}\right)-2|x-x'|^2\left(a{\partial^2 
\delta_1 \over (\partial t')^2}+b{\partial^2 \delta_2 \over 
(\partial t)^2}\right) \right]\eqno(13)$$
 In (13) the integrand is antisymmetric under 
transformations $$ t \to t',\ \ t'\to t,\ \ x\to x',\ \ x'\to x$$
if $$a=b \ \ (=1/2)$$.
 Then      the whole integral 
(13) has identically zero values:  
$$ \Delta P_{self}=0$$
 So for an extended in one dimension particle the 
total change in particle momentum due to its self-interaction is 
zero (if is taken the half-sum of retarded and advanced 
interactions).  Then the natural question arises - where is the 
source of radiated energy? - The answer is obvious and comes from 
eq. (9) - the source of radiated electromagnetic energy is the 
external force work. 

Similar result holds for extended in three dimension particle.

This conclusion we  formulate as the theorem:

{\it For extended charged objects (particles) internal relativistic 
electromagnetic forces, also as internal forces of another nature, 
give zero contribution to the total change of  momentum 
of these objects (particles).}

If to go further  and formulate the hypothesis:

{\it  Extended charged objects (particles) must have that form of 
charge density and current density, that leads  
to zero values of  total internal relativistic 
electromagnetic forces, also as to zero values of total internal 
forces of another nature.}

Then in classical electrodynamics there would be no need to use in 
eq.  of particle motion radiation reaction force in Lorentz-Dirac or 
another form.  \eject \centerline {\bf{REFERENCES}} 
  \begin{enumerate}
  \item
  F.Rohrlich, {\it Classical Charged Particles}, Addison-Wesley,
  Reading, Mass., 1965.
\item
A.Sokolov, I.Ternov, {\it Syncrotron Radiation}, Pergamon Press,
    NY,1968. A.Sokolov, I.Ternov, {\it Relativistic
Electron}, (in russian), Nauka, Moscow, 1983.
\item D.Ivanenko, A.Sokolov,  {\it Classical field theory}, (in
russian), GITTL, Moscow, 1949
 \item S.Parrott, {\it
Relativistic Electrodynamics and Differential Geometry},
Springer-Verlag, NY, 1987.

\item S.Parrott, Found.Phys., 23 (1993), 1093.
\item W.Troost et al.,  preprint
hep-th/9602066.
\item Alexander A.Vlasov, preprints hep-th/9702177; hep-th/9703001.
\item Anatolii A.Vlasov, {\it Statistical Distribution Functions}, 
(in russian), Nauka, Moscow, 1966.
\item Anatolii A.Vlasov, {\it Non-local Statistical Mechanics },  
(in russian), Nauka, Moscow, 1978. 
\item A.Lozada, J.Math.Phys., 30 (1989), 1713.
 \end{enumerate}

 \end{document}